\documentclass[pdflatex,sn-nature]{sn-jnl}

\usepackage{graphicx}%
\usepackage{multirow}%
\usepackage{amsmath,amssymb,amsfonts}%
\usepackage{amsthm}%
\usepackage{mathrsfs}%
\usepackage[title]{appendix}%
\usepackage{xcolor}%
\usepackage{textcomp}%
\usepackage{manyfoot}%
\usepackage{booktabs}%
\usepackage{algorithm}%
\usepackage{algorithmicx}%
\usepackage{algpseudocode}%
\usepackage{listings}%
\usepackage{braket}

\theoremstyle{thmstyleone}%
\theoremstyle{thmstyletwo}%

\theoremstyle{thmstylethree}%

\begin{document}

\title{
Multivariate unbounded quantum regression via log-ratio probabilities mitigating barren plateaus
}


\author*[1]{\fnm{Jaemin} \sur{Seo}}\email{jseo@cau.ac.kr}

\affil*[1]{\orgdiv{Department of Physics}, \orgname{Chung-Ang University}, \orgaddress{\city{Seoul}, \country{Republic of Korea}}}


\abstract{
Quantum neural networks (QNNs) have shown remarkable potential due to their capability of representing complex functions within exponentially large Hilbert spaces. However, their application to multivariate regression tasks has been limited, primarily due to inherent constraints of traditional approaches that rely on Pauli expectation values. In this work, we introduce a novel and simple post-processing method utilizing log-ratio probabilities (LRPs) of quantum states, enabling efficient and unbounded multivariate regression within existing QNN architectures. Our approach exponentially expands the number of regression outputs relative to qubit count, thus significantly improving parameter and qubit efficiency. Additionally, by enhancing parameter dependencies in the cost function and leveraging gradient pumping effects from the log-ratio transformation, our method mitigates the well-known barren plateau phenomenon, thereby stabilizing training. We further demonstrate that this approach facilitates robust uncertainty quantification, capturing both epistemic and aleatoric uncertainties simultaneously. Our findings underscore the practical potential of LRP-QNNs for complex multi-output regression tasks, particularly within current resource-constrained quantum hardware environments.
}



\maketitle


Recent advances in quantum technologies, such as creating and entangling qubits, have accelerated the development of quantum computing and quantum machine learning \cite{Biamonte2017_qml, Cerezo2022_qml}. A representative example is the quantum neural network (QNN), also known as parameterized quantum circuits (PQCs), which employ the angles of unitary transformations in quantum circuits as learnable parameters, enabling training as a black-box model \cite{Benedetti_2019_pqc, Hubregtsen2021_pqc, ieee2023_pqc}. A QNN composed of $n$ qubits can represent up to $2^n$ quantum states, thereby offering significantly higher expressivity in Hilbert space compared to classical neural networks (NNs) with the same number of parameters \cite{Benedetti_2019_pqc, Du_PRR2020_QNN_expressivity}. Moreover, since the measurement outcomes inherently yield probabilistic states depending on the parameters, QNNs naturally function as classifiers without the artificial introduction of softmax processes, making them particularly suited for classification tasks \cite{Senokosov_MLST_2024_image_probabilistic_classification}.

For regression tasks, on the other hand, QNNs typically utilize the expectation value of a Pauli operator, not using the probabilistic quantum states themselves. The Pauli operators serve as measurement observables in quantum circuits, projecting quantum states onto specific axes of the Bloch sphere to extract meaningful information such as expectation values. This approach provides continuous expectation values between $-1$ and $1$ through repeated measurements, which can then be rescaled to arbitrary upper and lower bounds, thus enabling regression over continuous values \cite{Qi2023_npjqi_pauli_single_output}.

However, the conventional approach of projecting exponentially expressive quantum states onto a single global parameter via Pauli operators introduces several issues not encountered by classical NNs. Firstly, it does not inherently support multivariate output prediction; it can only output at most as many features as the number of qubits, or requires multiple quantum circuits to perform multivariate regression \cite{Ding2024_SR_multi_circuit}. This is unfavorable for current hardware with few qubits. Secondly, the outputs inherently have bounded ranges, requiring prior knowledge of data distribution for appropriate rescaling. This limitation poses significant challenges for ``data-sparse'' learning tasks, such as few-shot extrapolation tasks or physics-informed neural networks \cite{RAISSI_2019_pinn, Seo_SR_2024_pinn}. Lastly, since the cost function is defined based on expectation values projected onto a single axis, parameter dependencies across the remaining $2^n-1$ dimensions diminish. This exacerbates the barren plateau phenomenon \cite{McClean2018_NC_original_barren_plateau}, a critical issue in QNN training characterized by flat loss landscapes across large parameter spaces. These three issues, nonexistent in classical NNs, must be resolved for QNNs to surpass the capabilities of classical NNs.

In this study, we propose a method leveraging the log-ratio probabilities of the possible output quantum states in QNNs, enabling efficient multivariate and unbounded regression. Additionally, by exploiting the enhanced parameter dependency of the cost function across multiple output states and gradient pumping effects arising from log-ratio transformations, our method mitigates gradient vanishing, thereby alleviating the barren plateau phenomenon. This approach requires minimal structural modification to existing QNN frameworks and is implemented via a straightforward post-processing step, making it readily applicable to current well-developed QNN techniques.

\section*{Results}\label{sec_results}

\subsection*{Quantum neural network with log-ratio probabilities}\label{sec2}

\begin{figure}[t]
\centering
\includegraphics[width=5.3in]{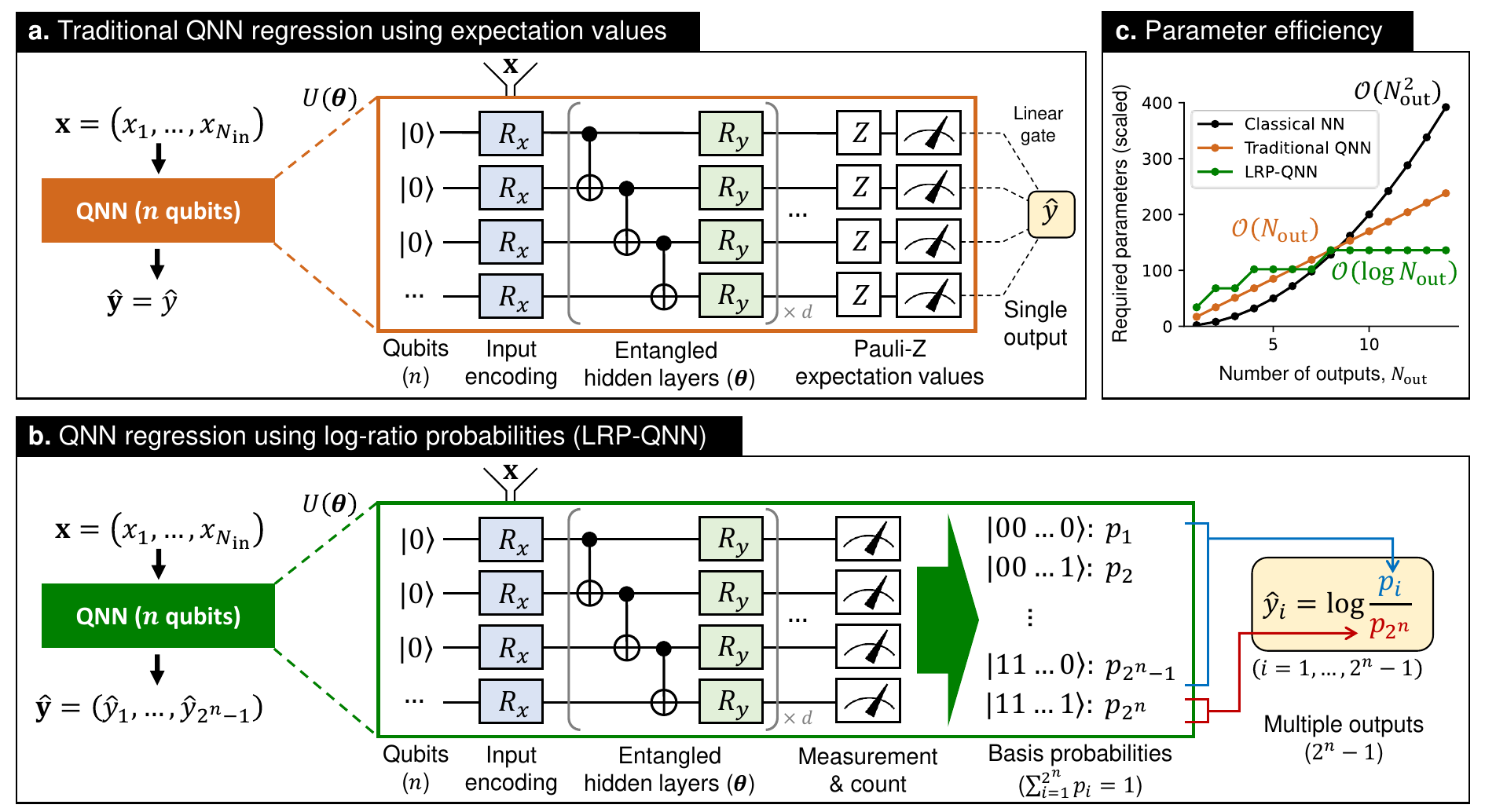}
\caption{\textbf{Architecture of QNN using log-ratio probabilities for multivariate regression.} \textbf{a}, Workflow of the traditional QNN using Pauli expectation values. \textbf{b}, Workflow of our proposed method, the QNN using log-ratio probabilities (LRP-QNN). \textbf{c}, Comparison of the parameter complexity among the classical NN, the traditional QNN, and the LRP-QNN.}
\label{fig1}
\end{figure}

With recent progress in QNNs, their applications have expanded beyond their natural role in classification to various regression tasks. A QNN consists of a quantum circuit with unitary rotations and entanglement layers among qubits. Typically, Pauli-Z ($Z$) expectation values are estimated through measurements after applying these parameterized layers, as shown in Figure \ref{fig1}a. Either a linear combination (e.g., average value) of the Pauli expectation values of each qubit or the expectation value of a single representative qubit can be used for the regression output. These expectation values take continuous values between $-1$ and $1$, corresponding to the eigenvalues of Pauli operators, and can be rescaled to perform regression over bounded expanded intervals.

This Pauli expectation value-based QNN is generally a structure suitable for ``single-output'' or ``few-output'' regression. In many practical problems, however, it is often necessary to predict multiple independent physical quantities \cite{Seo_NF2021, Shousha_2024} or generate high-dimensional outputs such as images, which may involve hundreds or more regression targets \cite{NEURIPS2021_diffusion_model}, which have been easily handled by classical NNs. When using QNNs, even if the expectation values of every qubit are considered as independent outputs, as many qubits as the required output targets are required, which is not very feasible in current few-qubit devices.

To address this gap, we propose a single post-processing method that enables: (a) multivariate output regression, (b) unbounded continuous output ranges, and (c) mitigation of barren plateaus. As illustrated in Figure \ref{fig1}b, an $n$-qubit QNN has $2^n$ possible states, and repeated measurements allow estimation of the probability of each state ($p_i$). While these probabilities take continuous values between 0 and 1 and could naively be used for regression, they are strongly dependent due to the normalization constraint $\sum_i p_i=1$. To decouple this dependency and remove output bounds, we use a simple post-processing method that sacrifices one of the probabilities (e.g., $p_{2^n}$ in this study) to compute $2^n-1$ log-ratio probabilities (LRPs), as shown in Figure \ref{fig1}b:

\begin{equation}
    \hat{y}_i = \log \left.\frac{p_i}{p_{2^n}}\right|_{\theta}, \quad  i=1, 2, \dots, 2^n-1.
\label{eq1} \\
\end{equation}

The resulting outputs $\hat{y}_i$ are mutually independent, unconstrained by any normalization condition, and can span the entire real domain $(-\infty, \infty)$. Importantly, even though qubits may be entangled, each basis state is classically distinct and not entangled with other basis states, allowing the $\hat{y}_i$'s to be treated as independent outputs, making this representation directly suitable for multivariate regression.

This framework also offers significant efficiency in terms of the number of trainable parameters. In classical NNs, performing multivariate regression with $N_{\text{out}}$ outputs typically requires an output layer and hidden layers of size proportional to $N_{\text{out}}$, resulting in a total number of parameters that scales as $\mathcal{O}(N_{\text{out}}^2)$. In contrast, multi-output QNNs based on Pauli expectation values require $N_{\text{out}}$ separate qubits and wires, leading to parameter counts that scale linearly with $N_{\text{out}}$. Our proposed method, based on log-ratio probabilities, can represent up to $2^n$ outputs with only $n$ qubits, implying that the number of required parameters (and qubits) scales as $\mathcal{O}(\log{N_{\text{out}}})$. As depicted in Figure \ref{fig1}c, this trend highlights the advantage of our method in multi-output scenarios such as high-resolution image generation, where efficient multivariate regression with a large number of outputs is crucial.

\subsection*{Mitigation of barren plateaus}\label{sec3}

\begin{figure}[t]
\centering
\includegraphics[width=5in]{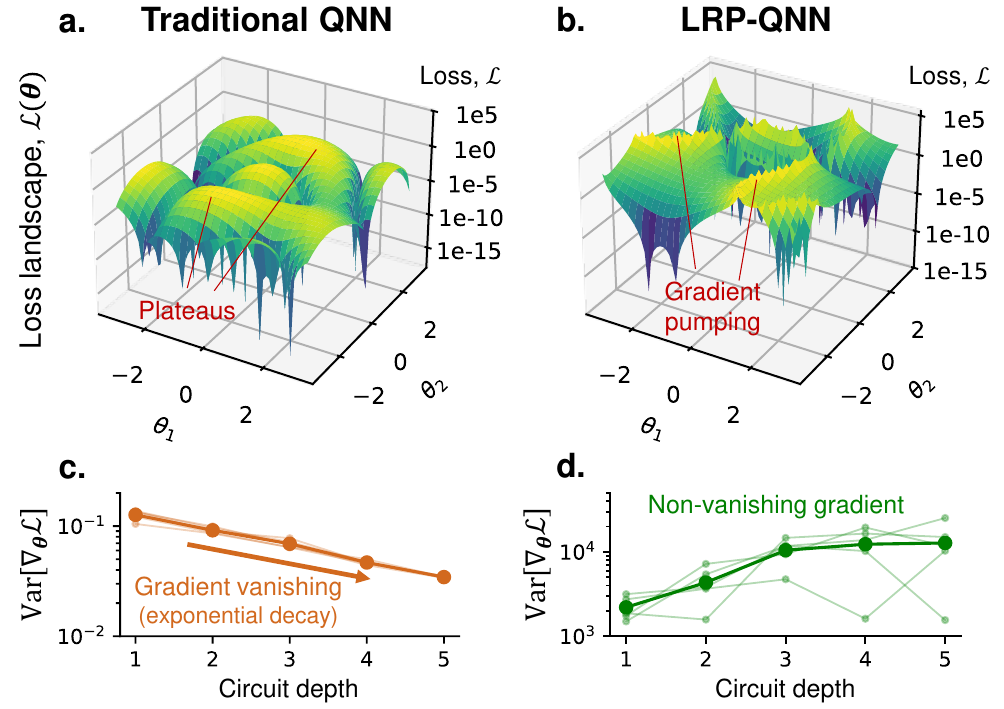}
\caption{\textbf{Comparison of barren plateaus in the traditional QNN and the LRP-QNN.} \textbf{a}, The loss landscape for the traditional QNN, showing several flat plateaus. \textbf{b}, The loss landscape for the LRP-QNN, showing steeper gradients and free of plateaus. \textbf{c}, Variance of the loss gradients in the traditional QNN, exponentially vanishing with respect to the circuit depth, which is major evidence of barren plateaus. \textbf{d}, Variance of the loss gradients in the LRP-QNN, which does not vanish with respect to the circuit depth.}
\label{fig2}
\end{figure}

Although QNNs have garnered significant interest due to their high expressivity in high-dimensional Hilbert spaces \cite{Benedetti_2019_pqc, Du_PRR2020_QNN_expressivity}, the entanglement-enabled compression of information \cite{IEEE2022_quantum_data_compression}, and the potential for efficient sampling \cite{Coyle2020_npjqi_sampling_efficiency}, several critical hurdles remain. One of the most important issues is the so-called barren plateau phenomenon \cite{McClean2018_NC_original_barren_plateau}, where gradients vanish exponentially as the depth of a quantum circuit increases when its parameters are randomly initialized. This is fundamentally different from the vanishing gradient problem in classical NNs. In QNNs, randomly initialized parameters tend to produce observables that concentrate near their expected values. As the circuit complexity increases, the gradient of cost functions based on these observables tends to vanish, a point which will be revisited in Figure \ref{fig2}.

Moreover, because standard QNNs typically rely on Pauli projections, they discard information along directions orthogonal to the projection axis. This exacerbates the vanishing gradient problem by eliminating sensitivity to a large portion of the parameter space. In addition, the presence of non-zero noise in noisy intermediate-scale quantum (NISQ) devices can further contribute to the emergence of barren plateaus \cite{Wang2021_NC_barren_plateau_cause2}.

The $n$-qubit quantum state generated by a parameterized quantum circuit $U$ can be expressed as

\begin{equation}
    \ket{\psi(\boldsymbol{\theta})} = U(\boldsymbol{\theta}) \ket{0}^{\otimes n},
\label{eq2} \\
\end{equation}

\noindent with the probabilities of the measurable state $\ket{i}$, $p_i(\boldsymbol{\theta}) = \lvert \braket{i|\psi(\boldsymbol{\theta})} \rvert^2$. The normalization constraint is given by $\sum_{i=1}^{2^n}{p_i(\boldsymbol{\theta})} = 1$. If we use the Pauli-Z expectation value ($\braket{Z}=\braket{\psi(\boldsymbol{\theta})|Z|\psi(\boldsymbol{\theta})}$) of a single qubit as the regression output, the corresponding mean squared error (MSE) loss is given by:

\begin{equation}
    \mathcal{L}(\boldsymbol{\theta}) = \left( \braket{Z} - y_{\text{true}} \right)^2 = \left( \braket{\psi(\boldsymbol{\theta})|Z|\psi(\boldsymbol{\theta})} - y_{\text{true}} \right)^2.
\label{eq3} \\
\end{equation}

\noindent where $y_{\text{true}}$ is the target value. Here, we assume a single target batch for simplicity. The gradient of this loss with respect to a parameter $\theta_j$ is:

\begin{equation}
    \frac{\partial \mathcal{L}}{\partial \theta_j} = 2 \left( \braket{Z} - y_{\text{true}} \right) \cdot \frac{\partial \braket{Z}}{\partial \theta_j}.
\label{eq4} \\
\end{equation}

Here, the Pauli-Z expectation value corresponds to a projection along a specific axis. When the quantum state is randomly initialized, the state amplitudes are likely to be distributed nearly symmetrically between $+1$ and $-1$, resulting in frequent cancellations in the expectation value ($\braket{Z}=\sum_{i=1}^{2^n} p_i(\boldsymbol{\theta})\cdot (\pm 1) \sim 0$). Consequently, the gradient ${\partial \braket{Z}}/{\partial \theta_j}$ in Equation (\ref{eq4}) becomes highly susceptible to vanishing under parameter perturbations. A major characteristic of barren plateaus is that the gradient variance decreases exponentially with the model complexity \cite{McClean2018_NC_original_barren_plateau}, which hinders the growth toward large models applicable to practical tasks.

This barren plateau phenomenon can be visually confirmed through a simple experiment on the loss landscape. Figure \ref{fig2}a shows the loss landscape of a 2-qubit, 4-layer quantum circuit projected onto the plane spanned by the top two Hessian eigenvectors of the parameter space ($\theta_1, \theta_2$) \cite{IEEE2020_hessian_loss_landscape}. Several plateau regions where the gradient vanishes ($\nabla_{\theta_j} \mathcal{L} \sim 0$) can be observed. Figure \ref{fig2}c shows the variation of the loss gradient, with respect to the depth of the circuit, for 5 different seeds, with each averaged over 100 ensembles. As the circuit complexity increases, the variance of the gradients decreases exponentially, consistently in different seeds, in agreement with typical observations of barren plateaus \cite{McClean2018_NC_original_barren_plateau}.

Recently, it has been shown that the barren plateaus depend on the form of the loss function itself \cite{Cerezo2021_NC_cost_dependent_barren_plateau}. However, in practice, it is often infeasible to alter the loss function arbitrarily for a given task. To address this, rather than modifying the loss function, we introduce a simple post-processing step at the output stage of the QNN, the log-ratio probability (LRP) transformation defined in Equation (\ref{eq1}), so that it indirectly influences the loss landscape and its gradients.

When we perform multivariate regression with the outputs $\textbf{y}=(y_1, \dots, y_{2^n-1})$ using the log-ratio probabilities defined in Equation (\ref{eq1}), the MSE loss takes the following form:

\begin{equation}
    \mathcal{L}(\boldsymbol{\theta}) = \frac{1}{2^n-1} \sum_{i=1}^{2^n-1} \left( \hat{y}_i(\boldsymbol{\theta}) - y_{i, \text{true}} \right)^2  = \frac{1}{2^n-1} \sum_{i=1}^{2^n-1} \left( \log \frac{p_i(\boldsymbol{\theta})}{p_{2^n}(\boldsymbol{\theta})} - y_{i, \text{true}} \right)^2.
\label{eq5} \\
\end{equation}

The gradient of this loss with respect to a parameter $\theta_j$ is then given by:

\begin{equation}
    \frac{\partial\mathcal{L}}{\partial \theta_j} = \frac{2}{2^n-1} \sum_{i=1}^{2^n-1} \left( \log \frac{p_i}{p_{2^n}} - y_{i, \text{true}} \right) \cdot \left( \frac{1}{p_i} \frac{\partial p_i}{\partial \theta_j} - \frac{1}{p_{2^n}} \frac{\partial p_{2^n}}{\partial \theta_j} \right).
\label{eq6} \\
\end{equation}

Unlike Equation (\ref{eq4}), which depends on the gradient of a single expectation value, the above expression is based on the gradient of measurement probabilities. Due to the normalization constraint $\sum_i p_i=1$, these probabilities cannot all simultaneously approach zero, which suppresses global cancellation effects in the gradient sum. Furthermore, the negative correlation of $p_i$ and $p_{2^n}$ makes it difficult for the last term in Equation (\ref{eq6}) to vanish. Notably, even when $\partial p_i/\partial \theta_j \sim \mathcal{O}(1/2^n)$ is small when we use many qubits, the derivative of the log function introduces a scaling term $1/p_i \sim \mathcal{O}(2^n)$, which again amplifies the gradient, so-called gradient pumping. As a result, the gradient magnitude remains stable even when individual basis probabilities are low.

Finally, in multivariate regression, this log-ratio formulation naturally couples different parameters to different output components through whole basis probability matching. Each regression target ($y_{i, \text{true}}$) guides each probability ($p_i$) and associated parameters in the circuit, through the $\left( \log (p_i/p_{2^n}) - y_{i, \text{true}} \right)$ term in Equation (\ref{eq6}). This contrasts with the Pauli expectation-based approach, in which much of the parameter dependency is lost through projection. Together, the normalization constraint, gradient pumping, and preservation of parameter sensitivity act synergistically to mitigate gradient vanishing, making the training of deep and wide QNNs more stable and efficient.

Figure \ref{fig2}b illustrates the loss landscape of a log-ratio probability transformed QNN (LRP-QNN) under the same 2-qubit, 4-layer circuit configuration used in Figure \ref{fig2}a. While the traditional QNN exhibits clear flat regions with vanishing gradients, the landscape in the LRP-based model is significantly steep and free of plateaus. Notably, in regions with large loss values, gradient pumping enhances the magnitude of the gradient, enabling faster convergence. Additionally, increasing the depth of the circuit does not degrade the gradient variance in this formulation, as shown in Figure \ref{fig2}d. Rather, it enhances the trainability through pumping the gradient variances. Therefore, without modifying the loss function or optimizer, the proposed log-ratio probability post-processing enables not only multivariate regression but also a practical mitigation of the long-standing barren plateau issue in QNNs.

\subsection*{Multivariate regression with log-ratio probabilities}\label{sec4}

\begin{figure}[t]
\centering
\includegraphics[width=5.3in]{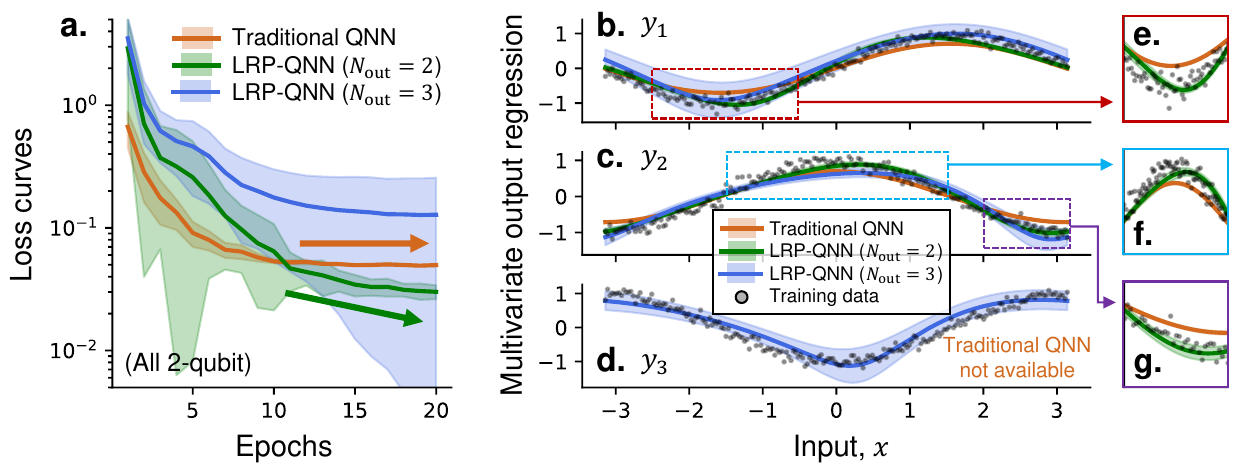}
\caption{\textbf{Multivariate regression capability of the QNNs.} \textbf{a}, The loss curves during the training of the traditional QNNs (orange) and the LRP-QNNs (green for two outputs and blue for three outputs). The shaded area indicates the variation among the 10 ensemble models with the same architecture. \textbf{b}, The first regression target of $y_1=\sin{x}$ with noise. \textbf{c}, The second regression target of $y_2=\cos{x}$ with noise. \textbf{d}, The third regression target of $y_3=-\cos{x}$ with noise. The traditional QNN with 2 qubits is unable to regress the third feature. \textbf{e-g}, The zoom-in plots of the regressions using the traditional QNN and the LRP-QNN.}
\label{fig3}
\end{figure}

To validate the multivariate regression capability and training efficiency of the proposed method, we applied it to an example multivariate output task as shown in Figure \ref{fig3}. This artificial task involves regressing three output features ($\sin x$, $\cos x$, and $-\cos x$) from a single input $x$. The ground truth training data, with added Gaussian noise $\mathcal{N}(0, 0.1^2)$ to demonstrate robustness, are indicated by black dots in Figures \ref{fig3}b-d. We further checked statistical coherence using an ensemble of 10 independent training runs, indicated with filled areas in Figure \ref{fig3}.

For comparison, we implemented both a traditional QNN (based on Pauli-Z expectation values) and an LRP-QNN (based on log-ratio probabilities), each using the same PQC structure with 2 qubits and 3 layers. The traditional QNN can output at most two features, corresponding to the expectation values of the individual two qubits. Therefore, for a fair comparison under identical capacity constraints, we first compare the performance of the two models on a 2-output ($\sin x$ and $\cos x$) regression task, colored with orange and green in Figure \ref{fig3}. As shown in Figure \ref{fig3}a, despite having identical architectural complexity, the loss reductions differ significantly. The traditional QNN (orange) produces bounded outputs, whereas the LRP-QNN (green) outputs unbounded values, resulting in larger initial loss values. However, the LRP-QNN exhibits a much faster loss reduction, probably due to the steep loss landscape shown in Figure \ref{fig2}b. Notably, the loss of traditional QNNs encounters a plateau after around 10 epochs, while the LRP-QNN loss continues to reduce and eventually reaches a lower loss. This result highlights that the simple addition of log-ratio probability post-processing can significantly improve the training efficiency of QNN in the same task, without modifying the underlying circuit structure. Interestingly, traditional QNN confidently makes incorrect predictions near its regression bound ($y_i=\pm1$), as shown in Figure \ref{fig3}e-g, whereas LRP-QNN does not show that problem since its regression is unbounded.

Furthermore, in this 2-qubit setup, the traditional QNN is inherently unable to regress the third feature ($-\cos x$ shown in Figure \ref{fig3}d), which underscores its inefficiency as a multivariate regressor. In contrast, our $n$-qubit LRP-QNN architecture enables access to $2^n-1$ output channels (3 in this 2-qubit example), offering exponentially scalable multivariate capability depending on the number of qubits. As the number of output features increases, the training complexity also increases, resulting in reduced training efficiency indicated by the blue curve in Figure \ref{fig3}a. Nonetheless, the LRP-QNN is still able to produce plausible estimates for a greater number of output features compared to the traditional QNN, as shown in Figures \ref{fig3}b–d.

\subsection*{Uncertainty quantification with LRP-QNNs}\label{sec5}

In the presence of noisy data, as observed in Figure \ref{fig3}, uncertainty quantification (UQ) becomes essential. A deterministic neural network, once trained, produces fixed predictions that cannot express any confidence interval or provide insight into the uncertainty or potential error of the prediction. Although epistemic uncertainty reflecting model uncertainty can be assessed by training multiple ensemble models with the same architecture, as in Figure \ref{fig3}, such an approach cannot adequately capture aleatoric uncertainty, which arises from inherent data noise. To evaluate aleatoric uncertainty, the model must learn and predict statistical properties of the data distribution, such as the mean and variance. In this section, we propose an application of the LRP-QNN that leverages its multivariate regression capability to output both the mean and variance of the target distribution, thereby enabling aleatoric uncertainty estimation as well.

UQ in QNNs has been studied to some extent, though less extensively than in classical NNs. The intrinsic probabilistic property of QNNs has been naturally used to generate non-deterministic predictions over different quantum states, primarily in discrete classification tasks \cite{Senokosov_MLST_2024_image_probabilistic_classification}. Another approach involves quantum conformal prediction, which provides confidence intervals \cite{Park_IEEE_2024_conformal}. Bayesian QNNs have also been explored, where the angular parameters of the quantum circuit are treated as probabilistic distributions, and posterior inference is performed over them \cite{Nguyen_IEEE_2022_bqnn, Duffield_MLST_2023_bqnn, Kim_IEEE_2023_bqnn}. In contrast, our approach introduces a more intuitive and simple UQ method that simultaneously captures both epistemic and aleatoric uncertainty by leveraging the multivariate output structure of the LRP-QNN through a deep ensemble framework \cite{NIPS2017_deep_ensemble, GANAIE2022_deep_ensemble}.

\begin{figure}[t]
\centering
\includegraphics[width=5in]{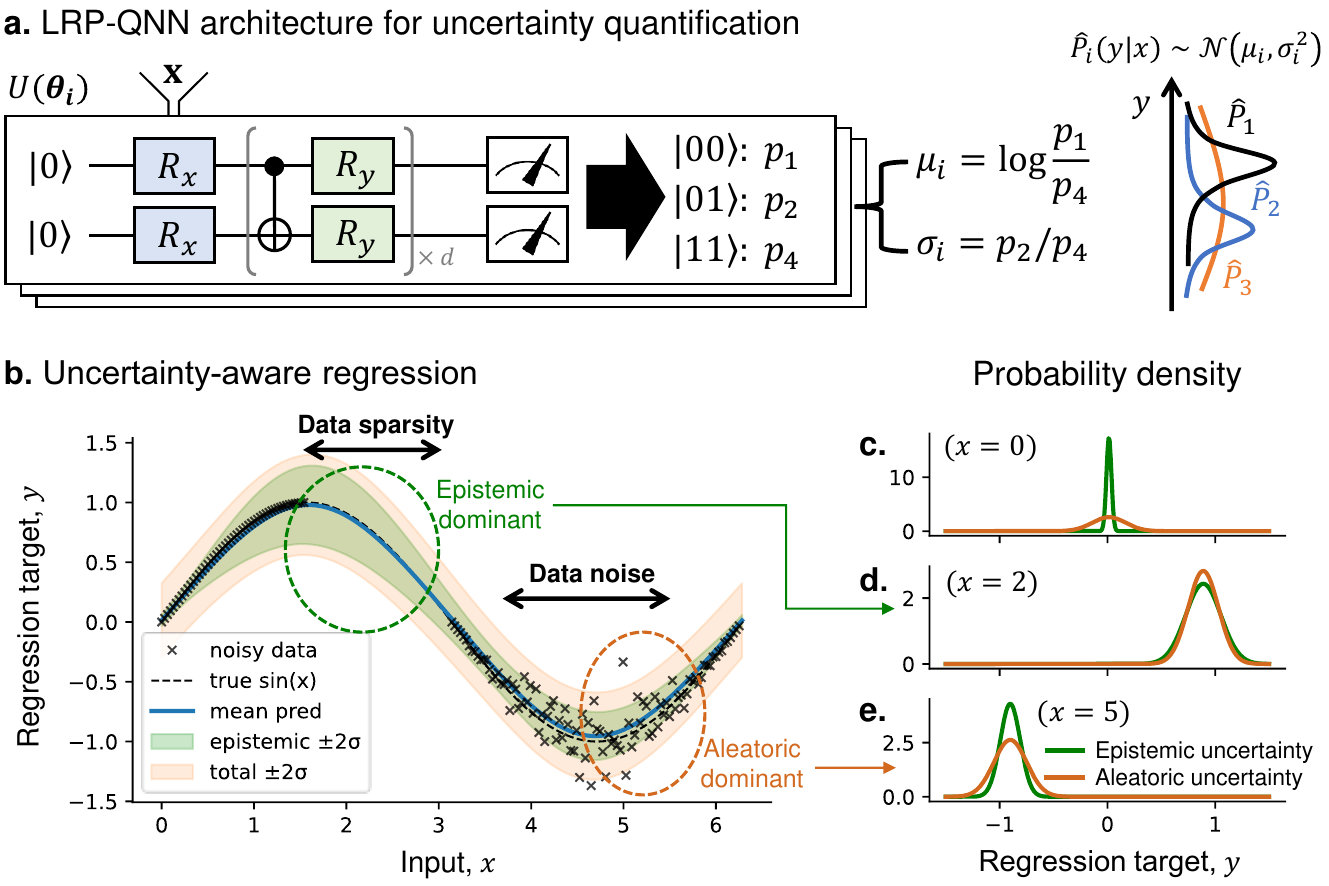}
\caption{\textbf{Deep ensemble-based uncertainty quantification using the multivariate capability of the LRP-QNN.} \textbf{a}, The workflow for the uncertainty quantification using the LRP-QNN, in the deep ensemble approach. \textbf{b}, The uncertainty-aware regression of the $y=\sin{x}$ dataset, where the data sparsity and injected noise are present. \textbf{c-e}, The probability density indicating the prediction and uncertainty quantification. The green curves represent the epistemic (model) uncertainty and the orange curves represent the aleatoric (data) uncertainty. The model estimates the epistemic uncertainty to be larger when the data has sparsity (\textbf{d}), and the aleatoric uncertainty to be larger when the data has noise (\textbf{e}).}
\label{fig4}
\end{figure}

As illustrated in Figure \ref{fig4}a, two of the outputs from a single 2-qubit LRP-QNN can be mapped to the predictive mean ($\mu_i$) and standard deviation ($\sigma_i$). The ensemble networks ($U(\boldsymbol{\theta}_i)$) are trained using the negative log-likelihood (NLL) loss, enabling the models to provide not only an estimate of the average prediction but also the aleatoric uncertainty, which reflects the spread of the data. By further utilizing the variance of $\mu_i$'s from the ensemble models, we can additionally capture epistemic uncertainty. The total uncertainty can then be quantitatively decomposed as:

\begin{equation}
    \text{Var}_{\text{tot}}(x) = \underbrace{\mathbb{E}[\sigma_i^2(x)]}_{\text{aleatoric}} + \underbrace{\text{Var}[\mu_i(x)]}_{\text{epistemic}}.
\label{eq7} \\
\end{equation}

Here, $\mathbb{E}[\cdot]$ and $\text{Var}[\cdot]$ are the mean and variance of the ensemble predictions for a given input $x$. To validate this UQ method, we construct an artificial $\sin{x}$ dataset in which both data sparsity (for epistemic uncertainty) and injected noise (for aleatoric uncertainty) are present, as shown by the black crosses in Figure \ref{fig4}b. The trained LRP-QNN not only predicts a rough estimate learned from the noisy data (the blue solid line), but also provides both epistemic and aleatoric uncertainty (the shaded areas). Moreover, we can observe that the model predicts larger epistemic uncertainty in regions where the data is sparse (Figure \ref{fig4}d), and larger aleatoric uncertainty in regions with noise in the data (Figure \ref{fig4}e).

Therefore, by utilizing the LRP-QNN's multivariate regression capability, we can construct a simple deep ensemble framework for quantifying uncertainty. This approach allows us not only to perform deterministic regression but also to estimate the full probability distribution of the outputs. As a future direction, we plan to extend this framework to mixture density prediction, enabling the model to represent distributions with multiple modes \cite{seo2025_qmdn}.

\section*{Discussions}\label{sec6}

This study proposes a quantum neural network (QNN) combined with a simple post-processing of log-ratio probabilities (LRPs), enabling simple and efficient multivariate unbounded regression, which has been challenging in conventional QNN approaches. To the best of our knowledge, we present the first demonstration of unbounded multivariate regression in QNNs without requiring circuit duplication or architectural modification. The traditional regression based on Pauli expectation values from an $n$-qubit QNN inherently limits the number of output features to $n$, with a bounded range. In contrast, our method utilizes log-ratios of the $2^n$ basis state probabilities generated from the quantum circuit, producing $2^n-1$ unbounded features, $\hat{y}_i=\log(p_i/p_{2^n})$. This leads to an exponential improvement in parameter and qubit efficiency. Furthermore, the LRP approach enhances parameter dependency in the loss function through multi-output matching of the basis probabilities, and induces gradient pumping due to the logarithmic form in the loss function. These effects effectively mitigate the barren plateau phenomenon, a long-standing barrier to the practical utilization of QNNs. Consequently, the LRP-based QNN can be effectively applied to various practical multi-output systems, and further enables uncertainty quantification with its multivariate capability.

Using LRP-QNNs for multivariate systems significantly enhances parameter and qubit efficiency, thus increasing usability within the current few-qubit and limited-memory quantum hardware. However, this approach introduces a trade-off in measurement complexity. The standard error of a single Pauli expectation value estimated from $N$ measurements follows $\mathcal{O}(1/\sqrt{N})$ due to its binomial nature. Thus, to achieve a desired accuracy $\epsilon$, the required number of measurements scales as $N_\text{exp}=\mathcal{O}(1/\epsilon^2)$. Extending this expectation-based approach to regress multiple output features ($M$ outputs) can require up to $N_\text{exp}=\mathcal{O}(M/\epsilon^2)$ measurements. In contrast, the LRP-based QNN output is defined as $\hat{y}_i=\log(p_i/p_{2^n})$. In multinomial sampling under the worst-case scenario (with balanced probabilities $p_i \approx 1/2^n$), the standard error for estimating each probability is $\sigma_i=\mathcal{O}(\sqrt{p_i(1-p_i)/N}) \approx \mathcal{O}(1/\sqrt{2^nN})$. Using the uncertainty propagation formula, the standard error of the output $\hat{y}_i$ becomes:

\begin{equation}
    \sigma_{\hat{y}_i}^2=\frac{\sigma_i^2}{p_i^2} + \frac{\sigma_{2^n}^2}{p_{2^n}^2} \approx \mathcal{O}(2^n/N),
\label{eq8} \\
\end{equation}

\noindent indicating that the number of measurements required scales as $N_\text{LRP}=\mathcal{O}(2^n/\epsilon^2)$. Considering that the maximum number of output features is $M = 2^n-1$, we have $N_\text{exp}<N_\text{LRP}$. This means that the LRP-based approach requires more measurements than the traditional QNN to achieve the same level of accuracy. Especially when Pauli observables commute and multiple output values can be simultaneously obtained from a single measurement, the expectation-based method yields $N_\text{exp}=\mathcal{O}(1/\epsilon^2) \ll N_\text{LRP}$.

Thus, the LRP-based QNN offers a significant trade-off: substantial parameter and qubit savings at the expense of increased computational time in the form of measurement repetitions. Nonetheless, this approach is expected to be a highly feasible and practical alternative, particularly in environments with few-qubit and limited-memory quantum hardware that permits frequent repeated measurements.


\section*{Methods}\label{sec7}

\subsection*{Parameterized quantum circuits}\label{subsec7.1}

Parameterized quantum circuits (PQCs) are quantum circuits composed of unitary rotations with tunable angle parameters \cite{Benedetti_2019_pqc, Hubregtsen2021_pqc, ieee2023_pqc}. These circuits are the building blocks of quantum neural networks (QNNs), which enable hybrid quantum-classical neural computing. A PQC consists of alternating gates of single-qubit rotations and cross-qubit entanglements:

\begin{equation}
    U(\boldsymbol{\theta})=\prod_{l=1}^L \left( U_{ent}^{(l)} \cdot U_{rot}^{(l)}(\boldsymbol{\theta}^{(l)}) \right)
\label{eq9} \\
\end{equation}

\noindent where $U_{rot}^{(l)}(\boldsymbol{\theta}^{(l)})$ is composed of single-qubit rotation gates ($R_x, R_y, R_z$) parameterized by $\boldsymbol{\theta}^{(l)}$ in the layer $l$. $U_{ent}^{(l)}$ applies fixed two-qubit entangling CNOT gates across the qubits. The angle parameters $\boldsymbol{\theta}^{(l)}$ at each layer $l$ consist of the rotating angles ($\phi_j^{(l)}, \theta_j^{(l)}, \omega_j^{(l)}$), where $j$ is the qubit index. A parameterized rotation gate on qubit $j$ is given by

\begin{equation} \label{eq10}
\begin{split}
    \left.U_{rot}^{(l)}(\boldsymbol{\theta}^{(l)})\right|_j &= R_z(\omega_j^{(l)})R_y(\theta_j^{(l)})R_z(\phi_j^{(l)}) \\
    &= 
    \begin{bmatrix}
    e^{-i(\phi_j^{(l)}+\omega_j^{(l)})/2} \cos(\theta_j^{(l)}/2) & \textbf{ } -e^{i(\phi_j^{(l)}-\omega_j^{(l)})/2} \sin(\theta_j^{(l)}/2) \\
    e^{-i(\phi_j^{(l)}-\omega_j^{(l)})/2} \sin(\theta_j^{(l)}/2) & e^{i(\phi_j^{(l)}+\omega_j^{(l)})/2} \cos(\theta_j^{(l)}/2) 
    \end{bmatrix}.
\end{split}
\end{equation}

The parameters $\boldsymbol{\theta}=\{\boldsymbol{\theta}^{(1)}, \dots, \boldsymbol{\theta}^{(L)}\}$ can be tuned via classical gradient descent algorithms based on a user-defined loss function. The expressivity of PQCs increases with the circuit complexity, enabling the representation of highly complex functions in exponentially large Hilbert spaces.

\subsection*{Numerical environments}\label{subsec7.2}

To design and construct the quantum circuits used in this study, we utilized the PennyLane library \cite{bergholm2022pennylane}, while the modeling of QNNs was performed using the PyTorch framework \cite{paszke2019pytorch}. The angle parameters within the circuit were trained using the Adam optimizer \cite{kingma2017adam}, with a learning rate set to 0.1 in all experimental examples presented in this work. For more efficient training, the input dataset was divided into mini-batches of size 64.

All QNN models used in the benchmark examples of Figures \ref{fig3} and \ref{fig4} share the same architecture of a 2-qubit, 3-layer PQC. The input variable $x$ is encoded via angle embedding using X-axis rotations, and within each layer, the quantum state is transformed using parameterized rotations (Equation (\ref{eq10})) and CNOT entanglement gates. At the final measurement stage, Pauli-Z expectation values are measured for the traditional QNN baselines, while the LRP-QNNs measure the probabilities of each basis state, which are then used in Equation (\ref{eq1}). To assess statistical consistency, we trained an ensemble of 10 models with identical architecture for each experiment. The Python codes that generate the dataset and train the QNNs can be seen at https://github.com/jaem-seo/quantum-multi-regression.

\subsection*{Related works}\label{subsec7.3}

In early uses of QNNs for regression, radial basis function (RBF)-based quantum networks were explored, although with limited performance \cite{Shao_PRA_2020_QRBF}. Alternatively, hybrid models were proposed in which quantum circuits were used only for embedding and feature extraction, while classical NNs were employed for the final regression output \cite{Reddy_2021_MLST_hybrid_regression}. Later, it was shown that even a single-qubit circuit could approximate any bounded function \cite{Perez_PRA_2021_one_qubit_universal}, leading to further studies on the maximal expressivity of QNNs in regression tasks \cite{Panadero2024_SR_qnn_maximal_expressivity}. While multivariate input regression using QNNs has been investigated \cite{Hirai2024_SR_multi_input}, the use of QNNs for multivariate output regression remains relatively unexplored. Some restricted efforts have been made to implement multi-output regression by associating each qubit with a Pauli projection \cite{Qi2023_npjqi_pauli_single_output}, but this approach suffers from limitations in delivering truly independent and unbounded outputs due to the inherent entanglement between qubits.

\section*{Acknowledgements}

This work was supported by a National Research Foundation of Korea (NRF) grant funded by the Korea government (MSIT) (Grant No. RS-2024-00346024).


\end{document}